\documentclass[12pt]{article}

\usepackage{amsfonts,amssymb,amsmath, hyperref}
\usepackage[final]{epsfig}
\usepackage{graphicx}

\date{}

\begin{document}

\title{ADELES IN MATHEMATICAL PHYSICS \\ {}  {}}
{}\bigskip {}
\bigskip
\bigskip
\bigskip
\author{Branko Dragovich\footnote{\textsf{\, E-mail:\,dragovich@phy.bg.ac.yu}} \\Institute of Physics, P.O. Box 57,
11001 Belgrade,
 Serbia }

\maketitle
\date{}

\bigskip
\bigskip
\bigskip

{\it \hskip4.5cm  Dedicated to Yakov Valentinovich Radyno}

{\it \hskip4.5cm  on the occasion of his 60th birthday}

\bigskip
\bigskip
%\bigskip

\begin{abstract}
Application of adeles in modern mathematical physics is briefly
reviewed. In particular, some adelic products are presented.
\end{abstract}

\section{Introduction}

$p$-Adic numbers are invented by K. Hansel in 1897. Ideles and
adeles are introduced by C. Chevalley and A. Weil, respectively, in
the 1930s. $p$-Adic numbers and adeles have many applications in
mathematics, e.g. representation theory, algebraic geometry and
modern number theory. Since 1987, $p$-adic numbers and adeles have
been used in construction of many models in modern mathematical
physics and related topics. Here we consider applications of adeles
in mathematical physics.

\section{Adeles}

On the field $\mathbb{Q}$ of rational numbers any non-trivial norm
is equivalent either to the usual absolute value $|\cdot|_\infty$ or
to a $p$-adic absolute value $|\cdot|_p$  (Ostrowski theorem). For a
rational number $x = p^\nu \, \frac{a}{b}$, where integers $a$ and
$b$ are not divisible by prime number $p$, by definition $p$-adic
absolute value is $|x|_p = p^{-\nu}$ and $|0|_p = 0 .$ This $p$-adic
norm is a non-Archimedean (ultrametric) one, because $|x + y|_p \leq
\mbox{max} \{|x|_p \,, |y|_p \}$. As completion of $\mathbb{Q}$
gives the field $\mathbb{Q}_\infty \equiv \mathbb{R}$ of real
numbers with respect to the $|\cdot|_\infty$, by the same procedure
one get the fields $\mathbb{Q}_p\,\,$ of $p$-adic numbers ($p = 2,\,
3\,, 5\, \cdots$) using $|\cdot|_p$ . Any number $x \in
\mathbb{Q}_p\,\,$ has a unique canonical representation
$$
x = p^{\nu (x)} \, \sum_{n = 0}^{+ \infty} \, x_n\, p^n  \,, \quad
\nu (x) \in \mathbb{Z}\,, \quad x_n \in \{0,\, 1,\, \cdots, \, p-1
\} . \eqno(1)
$$

Real and $p$-adic numbers, as completions of rationals, unify by
adeles. An adele $\alpha$ is an infinite sequence
$$
\alpha = (\alpha_\infty ,\, \alpha_2 ,\, \alpha_3, \, \cdots ,\,
\alpha_p \,,\, \cdots) \,,           \quad \alpha_\infty \in
\mathbb{R} \,, \,\, \alpha_p \in \mathbb{Q}_p \,,   \eqno(2)
$$
where for all but a finite set $\mathcal{P}$ of primes $p$ one has
that $\alpha_p \in \mathbb{Z}_p = \{ x\in \mathbb{Q}_p \, : |x|_p
\leq 1 \}$. $\mathbb{Z}_p$ is the ring of $p$-adic integers. The set
$\mathbb{A}_{\mathbb{Q}}$ of all adeles can be presented as
$$
\mathbb{A}_{\mathbb{Q}} = \bigcup_{\mathcal{P}} A (\mathcal{P})\,,
\quad A (\mathcal{P}) = \mathbb{R}\times \prod_{p\in \mathcal{P}}
\mathbb{Q}_p \times \prod_{p \not\in \mathcal{P}} \mathbb{Z}_p \,.
\eqno(3)
$$
Endowed with componentwise addition and multiplication
$\mathbb{A}_{\mathbb{Q}}$ is the adele ring.

The multiplicative group of ideles $\mathbb{A}_{\mathbb{Q}}^\times$
is a subset of $\mathbb{A}_{\mathbb{Q}}$ with elements $\eta = (\eta
_\infty\,, \eta _2\,, \eta _3 \,, \cdots , \eta _p\,, \cdots)$ ,
where $\eta _\infty \in \mathbb{R}^\times = {\mathbb R} \setminus \{
0\}$ and $\eta _p \in \mathbb{Q}^\times_p = {\mathbb Q}_p \setminus
\{0 \}$ with the restriction that for all but a finite set $\mathcal
P$  one has that $\eta _p \in {\mathbb U}_p = \{ x \in
\mathbb{Q}_p\, : |x|_p = 1 \}$ . Thus the whole set of ideles is
$$
 {\mathbb A}_{\mathbb{Q}}^\times = \bigcup_{{\mathcal P}} { A}^\times ({\mathcal P}),
 \ \ \ \  A^\times ({\mathcal P}) = {\mathbb R}^{\times}\times \prod_{p\in {\mathcal P}}
 {\mathbb Q}^\times_p
 \times \prod_{p\not\in {\mathcal P}} {\mathbb U}_p \, .
 \eqno(4)
$$

A principal adele (idele) is a sequence $ (x, x, \cdots, x, \cdots)
\in \mathbb{A}_{\mathbb{Q}}$ , where $x \in \mathbb{Q}\quad (x \in
{\mathbb Q}^\times )$. ${\mathbb Q}$ and ${\mathbb Q}^\times$ are
naturally embedded in $\mathbb{A}_{\mathbb{Q}}$ and
$\mathbb{A}_{\mathbb{Q}}^\times$ , respectively.

Let  ${\mathbb P}$ be set of all  primes $p$. Denote by
$\mathcal{P}_i \,, \,\, i \in \mathbb{N} ,$ subsets of ${\mathbb
P}$. Let us introduce an ordering by ${{\mathcal P}_i} \prec
{\mathcal P}_j$ if ${{\mathcal P}_i} \subset {\mathcal P}_j$. It is
evident that ${ A}({\cal P}_i)\subset { A}({\cal P}_j)$ when
${\mathcal P}_i \prec {\mathcal P}_j$. Adelic topology in
$\mathbb{A}_{\mathbb{Q}}$ is introduced by inductive limit: $
\mathbb{A}_{\mathbb{Q}} = \lim \mbox{ind}_{{\mathcal P}} { A}({\cal
P})$. A basis of adelic topology is a collection of open sets of the
form $ V ({\mathcal P}) = {\mathbb V}_\infty \times \prod_{p \in
{\mathcal P}} {\mathbb V}_p\, \times \prod_{p \not \in {\mathcal P}}
{\mathbb Z}_p \, $, where ${\mathbb V}_\infty$ and ${\mathbb V}_p$
are open sets in ${\mathbb R}$ and ${\mathbb Q}_p$ , respectively. A
sequence of adeles $\alpha^{(n)}\in {\mathbb A}_{\mathbb{Q}}$
converges to an adele $\alpha \in {\mathbb A}_{\mathbb{Q}}$ if
$({\it i})$ it converges to $\alpha$ componentwise and $({\it ii})\,
$ if there exist a positive integer $N$ and a set ${\mathcal P}$
such that $\, \alpha^{(n)}, \, \alpha \in  A ({\mathcal P})$ when
$n\geq N$. In the analogous way, these assertions hold also for
idelic spaces ${ A}^\times({\mathcal P})$ and $\mathbb{
A}_{\mathbb{Q}}^\times$. $\mathbb{ A}_{\mathbb{Q}}$ and ${\mathbb
A}_{\mathbb{Q}}^\times$ are locally compact topological spaces.

For various mathematical aspects of adeles one can see books [1, 2,
3].

\section{Adelic models}

Recall that results of measurements  are rational numbers, and
physical models have been treated using real and complex numbers.
Since $\mathbb{Q}$ is dense not only in $\mathbb{R}$ but also in
$\mathbb{Q}_p$, it has been natural to expect some applications of
$p$-adic numbers in mathematical modeling of physical systems. First
significant employment of $p$-adic numbers in physics started in
1987 by successful construction of $p$-adic string amplitudes. From
the very beginning there has been an opinion that all prime numbers
should be equally important and that $p$-adic models  should be
somehow connected with standard ones (over real or complex numbers).
According to the Hasse local-global principle an equation has a
solution over $\mathbb{Q}$ if and only if it has solutions over
$\mathbb{R}$ and all $\mathbb{Q}_p$. These ideas naturally gave rise
to an application of adeles and construction of adelic physical
models (for an early review, see [4, 5]).

Especially so-called adelic products have been attracted much
attention. They are of the form
$$
\phi_\infty (x_1\,,\cdots\,, x_n\,;  a_1\,, \cdots\,, a_m)
\prod_{p\in \mathbb{P}} \phi_p (x_1\,,\cdots\,, x_n\,;  a_1\,,
\cdots\,, a_m) = C  \,, \eqno(5)
$$
where $\phi_\infty$ and $\phi_p$ are real or complex valued
functions, $x_i \in \mathbb{Q}\,, \quad a_j \in \mathbb{C}\,, \,\,
$ and $C$ is a constant (often $C = 1$). It is obvious that
expressions of the form (5) connect real and $p$-adic
characteristics of the same object at the equal footing. Moreover,
the real quantity $\phi_\infty (x_1\,,\cdots\,, x_n\,; a_1\,,
\cdots\,, a_m)$ can be expressed as product of all $p$-adic
inverses. This can be of practical importance when functions
$\phi_p$ are simpler than $\phi_\infty$, but may also lead to more
profound understanding of physical reality.

For the reason of better understanding, let us first present two
simple examples:
$$
|x|_\infty \times \prod_{p\in \mathbb{P}} |x|_p = 1\,, \,\,
\mbox{if} \,\, x \in \mathbb{Q}^\times \,, \quad \mbox{and} \quad
\chi_\infty (x) \times \prod_{p\in \mathbb{P}} \chi_p (x) = 1 \,,
\,\, \mbox{if} \,\, x \in \mathbb{Q} \,, \eqno(6)
$$
where $\chi_\infty (x) = \exp ( -{2 \pi i x})$ and $\chi_p (x) =
\exp 2 \pi i \{x\}_p$ are real and $p$-adic additive characters,
respectively, and $\{x\}_p$ denotes the fractional part of $x$. It
follows from (6) that $d_\infty (x, y) = \prod_{p\in \mathbb{P}}
d_p^{-1} (x, y)$, where $d_\infty (x, y) = |x - y|_\infty$ and $d_p
(x, y) = |x -y|_p$, i.e. the usual distance between any two rational
points can be regarded through product of the inverse $p$-adic ones.
One can also write $\chi_\infty (ax + b t)\, = \, \prod_{p\in
\mathbb{P}}\, \chi_p [-(ax + b t)]$ when $a\,, b\,, x\,, t \in
\mathbb{Q}$, and consider a real plane wave as composed of $p$-adic
plane waves.

Let us also notice some adelic products related to number theory:
$$
\lambda_\infty (x) \, \prod_{p\in \mathbb{P}} \lambda_p (x) = 1\,,
\quad \quad \quad \Big(\frac{x, y}{\infty} \Big) \, \prod_{p\in
\mathbb{P}} \Big(\frac{x, y}{p} \Big) = 1   \,, \eqno(7)
$$
where $x$ is presented by (1) and
$$
\lambda_p (x) = \left\lbrace
\begin{array}{rllll}
    & 1,   \hskip4.5cm  \nu (x) = 2 k \,, \quad p \neq 2\,,  \\
    & \sqrt{\Big( \frac{-1}{p} \Big)}\, \Big(\frac{ x_0}{p} \Big)\,, \hskip2cm \,\,\,\,\,\, \nu (x)
    = 2 k + 1\,, \quad p \neq 2 \,,  \\
    & \exp{[\pi i (x_1 + 1/4)]} \,, \hskip1.5cm \,\, \nu (x) = 2 k\,, \quad  p = 2\,, \\
    & \exp{[\pi i (x_2 + x_1/2 + 1/4)]} \,, \quad \nu (x) = 2 k + 1\,, \quad  p = 2\,, \\
\end{array}
\right. \eqno(8)
$$
$$
\lambda_\infty (x) = \exp{\Big(- \frac{\pi i}{4}\, \mbox{sgn} \,
x\Big)} \,, \quad \Big(\frac{x, y}{\infty} \Big) = \left\lbrace
\begin{array}{rll}
    - 1,& \quad \quad x < 0, \,\,\, y < 0 \,, \\
    1,& \quad \quad   $otherwise$\,,\\
\end{array}
\right. \eqno(9)
$$
 $\Big(\frac{x}{p}\Big)$ and $\Big(\frac{x, y}{p}\Big)$ are
Legendre and Hilbert symbols [5], respectively.

Gauss integrals satisfy adelic product formula [6]
$$
\int_{\mathbb{R}} \chi_\infty (a \, x^2 + b\, x)\, d_\infty x \,
\prod_{p\in \mathbb{P}} \int_{\mathbb{Q}_p} \chi_p (a \, x^2 + b\,
x)\, d_p x = 1 \,,   \quad a \in {\mathbb{Q}^\times}\,, \,\,\, b \in
{\mathbb{Q}} \,, \eqno(10)
$$
what follows from
$$
\int_{\mathbb{Q}_v} \chi_v (a \, x^2 + b\, x)\, d_v x  = \lambda_v
(a)\, |2 \,a|_v^{-\frac{1}{2}}\, \chi_v \Big(- \frac{b^2}{4 a}
\Big)\,, \quad v = \infty\,, 2\,, \cdots\,, p \cdots \,.  \eqno(11)
$$
These Gauss integrals apply in evaluation of the Feynman path
integrals
$$
\mathcal{K}_v (x'', t''; x',t') = \int_{x',t'}^{x'',t''} \chi_v
\Big(- \frac{1}{h}\, \int_{t'}^{t''} L (\dot{q}, q, t) \, dt \Big)\,
\mathcal{D}_v q  \,, \eqno(12)
$$
for kernels $\mathcal{K}_v (x'', t'' ; x', t')$ of the evolution
operator
 in adelic quantum mechanics [7]  for quadratic Lagrangians. In the case of Lagrangian
 $L (\dot{q}, q) =\frac{1}{2}\Big( - \frac{\dot{q}^2}{4} - \lambda\, q + 1 \Big)  $
 for  the de Sitter cosmological model (what is similar to a particle with constant
 acceleration $\lambda$) one obtains [8, 9]
 $$
\mathcal{K}_\infty (x'', T ; x', 0) \prod_{p\in \mathbb{P}}
\mathcal{K}_p (x'', T ; x', 0) = 1\,,  \quad x'',x', \lambda \in
\mathbb{Q} \,, \,\,\, T \in \mathbb{Q}^\times \,, \eqno(13)
 $$
where
$$
\mathcal{K}_v (x'', T ; x', 0) = \lambda_v (- 8 T)\, |4
T|_v^{-\frac{1}{2}}\, \chi_v \Big( -\frac{\lambda^2 \, T^3}{24} +
[\lambda \,(x'' + x') - 2]\frac{T}{4} + \frac{(x'' - x')^2}{8 T}
\Big)\,. \eqno(14)
$$
The  adelic wave function for the simplest ground state has the form
$$
\psi_{\mathbb{A}} (x) = \psi_\infty (x) \prod_{p\in \mathbb{P}}
\Omega (|x|_p) =  \left\lbrace
\begin{array}{rll}
   &\psi_\infty (x) ,& \,\,x \in \mathbb{Z},\\
   & 0,& \,\, x \in\mathbb{Q}\setminus \mathbb{Z}\,,\\
\end{array}
\right.  \eqno(15)
$$ where $\Omega (|x|_p) = 1 $ if  $ |x|_p \leq 1$ and  $\Omega (|x|_p)
= 0 $ if  $ |x|_p > 1$. Since this wave function is non-zero only in
integer points it can be interpreted as discreteness of the space
due to $p$-adic effects in adelic approach.

The Gel'fand-Graev-Tate gamma and beta functions [4, 5] are:
$$\Gamma_\infty (a) = \int_{\mathbb{R}} |x|_\infty^{a-1} \chi_\infty (x)\, d_\infty x =
\frac{\zeta (1 - a)}{\zeta (a)} \,, $$  $$  \Gamma_p\, (a) =
\int_{\mathbb{Q}_p} |x|_p^{a-1} \chi_p (x)\, d_p x = \frac{1 -
p^{a - 1}}{1 - p^{-a}} \,, \eqno(16)$$
$$
B_\infty (a, b) = \int_{\mathbb{R}} |x|_\infty^{a-1}\, |1
-x|_\infty^{b-1}\, d_\infty x = \Gamma_\infty (a)\, \Gamma_\infty
(b)\, \Gamma_\infty (c)\,, \eqno(17) $$
$$ B_p (a, b) =
\int_{\mathbb{Q}_p} |x|_p^{a-1}\, |1 -x|_p^{b-1}\, d_p x = \Gamma_p
(a)\, \Gamma_p (b)\, \Gamma_p (c)\,,  \eqno(18)
$$
where $a, b, c \in \mathbb{C}$ with condition  $a + b + c = 1$ and
$\zeta (a)$ is the Riemann zeta function. With a regularization of
the product of $p$-adic gamma functions one has adelic products:
$$
\Gamma_\infty (u) \prod_{p \in \mathbb{P}} \Gamma_p (u) = 1  \,,
 \quad  B_\infty (a, b) \prod_{p \in
\mathbb{P}} B_p (a, b) = 1  \,, \quad u\neq 0, 1\,, \quad u = a ,
b , c\,,   \eqno(19)
$$
where $ a+b+c = 1$. It is worth noting now that $B_\infty (a, b)$
and $B_p (a, b)$ are the crossing symmetric standard and $p$-adic
Veneziano amplitudes for scattering of two open tachyon strings.
There are generalizations of the above product formulas for
integration on quadratic extensions of $\mathbb{R}$ and
$\mathbb{Q}_p$, as well as on algebraic number fields, and they
include  scattering of closed strings [4, 10].

Introducing real, $p$-adic and adelic zeta functions as
$$
\zeta_\infty (a) = \int_{\mathbb{R}} \exp{(-\pi \,x^2)} \,
|x|_\infty^{a -1} \, d_\infty x = \pi^{-\frac{a}{2}} \, \Gamma \Big(
\frac{a}{2}\Big)\,, \eqno(20)
$$
$$
\zeta_p (a) = \frac{1}{1 - p^{-1}}\, \int_{\mathbb{Q}_p} \Omega
(|x|_p)\, |x|_p^{a-1}\, d_p x = \frac{1}{1 - p^{- a}} \,, \quad
\mbox{Re}\, a > 1 \,, \eqno(21)
$$
$$
\zeta_{\mathbb{A}} (a) = \zeta_\infty (a) \prod_{p\in \mathbb{P}}
\zeta_p (a)  = \zeta_\infty (a)  \zeta (a) \,, \eqno(22)
$$
one obtains
$$
\zeta_{\mathbb{A}} (1 - a)  = \zeta_{\mathbb{A}} (a)\,, \eqno(23)
$$
 where $\zeta_{\mathbb{A}} (a)$ can be called adelic zeta function.
Let us note that $\exp{(-\pi \,x^2)}$ and $\Omega (|x|_p)$ are
analogous functions in real and $p$-adic cases. Adelic harmonic
oscillator [7]  has connection with the Riemann zeta function.
Namely, the simplest vacuum state of the adelic harmonic oscillator
is the following Schwartz-Bruhat function:
$$
\psi_{\mathbb{A}} (x) = 2^{\frac{1}{4}}\, e^{- \pi \, x_\infty^2}
\,\prod_{p\in \mathbb{P}} \Omega (|x_p |_p)\,, \eqno(24)
$$
whose the Fourier transform  $$ \psi_{\mathbb{A}} (k) = \int
\chi_{\mathbb{A}} (k\, x)\, \psi_{\mathbb{A}} ( x ) =
2^{\frac{1}{4}}\, e^{- \pi \, k_\infty^2} \,\prod_{p\in \mathbb{P}}
\Omega (|k_p |_p)     \eqno(25)
$$ has the same form as  $\psi_{\mathbb{A}} (x)$. The Mellin
transform of $\psi_{\mathbb{A}} (x)$ is
$$
\Phi_{\mathbb{A}} (a) = \int \psi_{\mathbb{A}} (x)\, |x|^{a} \,
d_{\mathbb{A}}^\times x $$  $$ = \int_{\mathbb{R}} \psi_{\infty}
(x)\, |x|^{a - 1} d_\infty x \prod_{p\in \mathbb{P}} \,
 \frac{1}{1 - p^{-1}} \, \int_{\mathbb{Q}_p} \Omega (|x|_p) |x|^{a
-1}\, d_p x = \sqrt{2} \, \Gamma \Big( \frac{a}{2} \Big) \,
\pi^{-\frac{a}{2}} \, \zeta (a)   \eqno(26)
$$
and the same for $\psi_{\mathbb{A}} (k)$. Then according to the Tate
formula one obtains (23). It is remarkable that such simple physical
system as harmonic oscillator is related to so significant
mathematical object as the Riemann zeta function.

Recently [11] adelic properties of dynamical systems, which
evolution is governed by linear fractional transformations
$$
f (x) = \frac{a x + b}{ c x + d} \,, \quad  a, b, c, d, \in
\mathbb{Q}\,, \quad a d - b c = 1     \eqno(27)
$$ is investigated. It is shown that rational fixed points are
$p$-adic indifferent for all but a finite set $\mathcal{P}$ of
primes, i.e. only for finite number of $p$-adic cases a rational
fixed point may be attractive or repelling.

\section{Concluding remarks}

We presented a brief review of some important applications of adeles
in modern mathematical physics. We considered above simple cases of
adeles $\mathbb{A}_{\mathbb{Q}}$ consisting of completions of
$\mathbb{Q}$. There is also ring of adeles $\mathbb{A}_{\mathbb{K}}$
related to the completions of any global field ${\mathbb{K}}$. There
is a straightforward generalization of $\mathbb{A}_{\mathbb{Q}}$ to
the n-dimensional vector space $\mathbb{A}_{\mathbb{Q}}^n  =
\prod_{i=1}^n \mathbb{A}_{\mathbb{Q}}^{(i)}$ (see, e.g. [1]). Adelic
algebraic group $G (\mathbb{A}_{\mathbb{K}})$ is an adelization of a
linear algebraic group $G$ over completion fields $\mathbb{K}_v$ of
a global field $\mathbb{K}$ [1, 2, 3].

For a more detail insight into this  attractive and promising field
of investigations let us also mention a few additional topics.
Adelic quantum cosmology (for a review, see [9]) is an application
of adelic quantum mechanics [7] to explore very early evolution of
the universe as a whole. Adelic path integral [12] is a suitable
extension of the standard Feynman path integral and serves to
describe quantum evolution of adelic objects. Conjecture on the
adelic universe with real and $p$-adic worlds, as well as $p$-adic
origin of dark matter and dark energy are discussed in [9]. Adelic
summability [13] of perturbation series is an approach to summation
of divergent series in the real case when they are convergent in all
$p$-adic cases. Use of effective Lagrangians on real numbers for
$p$-adic strings has been very efficient in their application to
string theory and cosmology. Paper [14] is an attempt towards
effective Lagrangian for adelic strings without tachyons. Further
development of adelic analysis and, in particular, adelic
generalized functions [6, 15, 16] is one of mathematical
opportunities.

One can conclude that there has been a successful application of
adeles in modern mathematical physics and  that one can expect a
growing interest in their further mathematical developments as well
as in applications.

\bigskip

{\noindent \bf Acknowledgements}.  The work on this article was
partially supported by the Ministry of Science and Environmental
Protection, Serbia, under contract No 144032D.

\end{document}